\documentstyle[12pt]{article}
\def\prd#1{{\em Phys.~Rev.}~{\bf D#1},\ }
\def\prl#1{{\em Phys.~Rev.~Lett.}~{\bf #1},\ }
\def\plett#1{{\em Phys.~Lett.}~{\bf #1B},\ }
\def\np#1{{\em Nucl.~Phys.}~{\bf B#1},\ }
\def\deg{\ifmmode{^{\circ}}\else ${^{\circ}}$\fi}
\def\pri{^{\,\prime}}

\newcommand{\bi}{\begin{itemize}}
\newcommand{\ei}{\end{itemize}}
\def\ed{\end{document}}
\def\be{\begin{equation}}
\def\ee{\end{equation}}
\def\vev#1{\left<{#1}\right >}
\def\cond{\vev{\lambda\lambda}}
\def\lc{\Lambda_c}
\def\mp{m_{Pl}}
\def\lm{\lambda\lambda}
\def\half{\frac{1}{2}}
\def\quarter{\frac{1}{4}}
\def\thalf{{\textstyle\frac{1}{2}}}
\def\tquarter{{\textstyle\frac{1}{4}}}
\def\tthird{{\textstyle\frac{1}{3}}}
\def\tsixth{{\textstyle\frac{1}{6}}}
\def\tin{T_i}
\def\lcmp{\left(\frac{\lc}{\mp}\right)^{\half}}
\def\lcmpl{\left(\lc/\mp\right)^{\half}}
\def\drr{\left(\frac{\delta\rho}{\rho}\right)_{hor,\lambda}}
\def\nl{N_{\lambda}}
\def\tev{\ \mbox{TeV}}
\def\pri{^{\, \prime}}
\def\prii{^{\,\prime \prime}}
\def\ms{m_{SUSY}}
\def\msigp{m_{\sigma\pri}}
\def\gsigp{\Gamma_{\sigma\pri}}
\def\phif{\phi_{50}}
\def\etab{\left(\frac{b}{\eta^3}\right)^{\quarter}}
\def\ac{\left(\frac{a}{c^2}\right)^{\quarter}}
\begin{document}
\begin{titlepage}
\begin{flushright}
{\sl NUB-3082/93-Th}\\
{\sl November 1993}\\
hep-ph/9312282
\end{flushright}
\vskip 0.5in
\begin{center}
{\Large \bf Primordial Fluctuations from Gaugino Condensation}\\[.5in]
{Haim Goldberg}
\\[.1in]
{\sl Department of Physics}\\
{\sl Northeastern University}\\
{\sl Boston, MA 02115}
\end{center}
\vskip 0.4in
\begin{abstract}
I present a model for inflation based on the gaugino-dilaton dynamics of
supersymmetry breaking.
The inflaton is the dimension--1 scalar field $\phi$ related
to the gaugino condensate via $\lambda^T\lambda=\phi^3.$ Recent work
in this area is used to obtain two significant results:  (1) Scalar density
fluctuations at second horizon crossing are generated on scale
$\lambda$ with amplitude
\[
\drr=A\cdot \ 10^3\cdot\ \left(\frac{\ms}{\mp}\right)^{\half}\
\left(\frac{\lambda}{100\ \mbox{Mpc}}\right)^{0.03}\ \ ,
\]
where $A$ is a constant which depends on (unknown) details of
gaugino-dilaton dynamics in the strong-coupling phase. Agreement with COBE
results is obtained if $A\simeq 2.$ \ (2) Due to mixing with hidden sector
glueballs, the dilaton mass is large $(\sim \left(\mp\ \ms^2\right)^
{\frac{1}{3}}\sim 10^8\ \mbox{GeV}),$ and reheating takes place at
$T\sim 1\tev.$ This {\em necessitates} that the presently observed baryon
asymmetry be  generated at a cosmic temperature below 1 TeV.
\end{abstract}
\end{titlepage}
\setcounter{page}{2}
\noindent{\large\bf Introduction}\medskip

The inflationary paradigm \cite{guth} was invented as  a remedy for some of the
important shortcomings of the standard Friedman-Robertson-Walker
cosmology: an
early period  of exponential (or near-exponential) expansion of the cosmic
scale factor could account for the presently observed large entropy,
near-flatness and
near-homogeneity of the visible universe. As a great bonus, quantum
field
fluctuations in the inflationary de Sitter phase get frozen in as scalar
metric perturbations \cite{drr}. These in turn can act as seeds for future
structure formation, and can be detected via the
Sachs-Wolfe effect \cite{sachs} as
the recently observed \cite{cobe}
temperature fluctuations in the cosmic background
radiation (CBR).

Generic to models which successfully comply with the requirements of the
inflation \cite{steinharturner} is the presence of a scalar field (the
`inflaton') which rolls slowly toward a global minimum of the
potential \cite{models}.
Although scalar fields are ubiquitous in particle physics, the usual ones are
unsuitable candidates for the inflaton: the flatness of the potential required
to satisfy both the slow roll and density fluctuation requirements impose
conditions of extremely weak self-coupling on the scalar field.
Typically, a
quartic self-coupling is forced to be of order $10^{-12}$--$10^{-14},$ which
is difficult to assign to any of the standard scalar fields in the particle
physics gallery. So a continuing question with respect to inflation is, as
it has been for ten years, ``what is the inflaton?''

In this work, I would like to present the outlines of an inflationary
mechanism, in which the physics derives from the superstring- and
supergravity-based gaugino-condensate  model of supersymmetry breaking
\cite{nilles1,taylor,font,nilles2,binetruy}.
There will be no small coupling constant in the model: its role will be filled
by the small ratio $(\ms/\mp)$ \cite{nat}. In addition, one of the major
cosmological problems inherent in `hidden sector' models, the undesirable
persistence of very weakly coupled TeV relics \cite{tevrelics}, will be
addressed and
(hopefully!) shown to be a non-problem. It should be immediately noted that
the
inflaton in the present work is {\em not} the dilaton field $S$;
the problems associated with this possibility have been recently
discussed \cite{steinbru}. A (non-perturbative) stabilization of the dilaton
potential \cite{dine} will, however,  certainly be necessary, as will be seen
in the later discussion of relics.

I begin by presenting the components of the gaugino-dilaton model of
$SUSY$-breaking  necessary for the present work. An important
aspect (especially for the decay of relics)
is that a dilaton superfield $S$  is universally coupled to all
(hidden and unhidden) gauge supermultiplets $w_i^{\alpha}$ in a
contribution
to the superpotential
\be
W= S\sum_iw_i^{\alpha}w_{i\alpha}
\label{eq:w}
\ee
such that $\vev{S}=S_0=1/g^2.$ Then, as is well-known,
supersymmetry is broken if $F_S,$ the auxiliary field of $S,$  has a
non-zero vacuum expectation value. Since, via the equations of motion,
$F_S\sim \sum \lambda_i^T\lambda_i,$ supersymmetry will be broken if a
gaugino condensate forms.

In general, the combined gaugino-dilaton dynamics can be rather complex.
It can involve moduli $(T)$ as well as the dilaton, and has been discussed by
various
authors \cite{taylor}-\cite{binetruy}.
One approach which has been used is to `freeze' the
hidden sector gaugino-condensate $\cond\equiv \lc^3$ at some value, and
examine
the $S,T$ dynamics with $\cond$ as background \cite{font,nilles2,binetruy}.
Other authors \cite{taylor,ross}
have
considered the combined effective potential as a function of all three
fields, in order to see how the condensate itself may form. The approach
used
in the present work is most closely tied to that used in the latter two
references.
\bigskip

\noindent{\large\bf The Dynamical Model}\medskip

In order to make use of the gaugino-dilaton model for inflationary dynamics,
one needs to incorporate  the relevant fields and their  effective potential
into a time dependent framework. Following the work of Refs.~\cite{veneziano}
and \cite{taylor} I make the crucial assumption  that in the
strong-coupling phase,
the $\lm$ composite is represented by a dimension-3 operator in the
form of the cube of a scalar field, $\phi^3,$ with the effective kinetic energy
\be
{\cal L}_{kin}=\thalf \eta\dot\phi^2\ \ .
\label{eq:kineta}
\ee
A form of the effective potential leading to the
formation of the condensate in the strongly-interacting phase has recently
been
presented by de la Macorra and Ross \cite{ross}
paralleling the original dynamics proposed by
Nambu and Jona-Lasinio \cite{njl} -- namely, a destabilization of the
symmetric phase
by
quantum loop effects. A similar destabilization due to the effects of
moduli was found in \cite{taylor}. The
two salient features observed in Ref.~\cite{ross}
are (1) the energy density gap between the true vacuum and the perturbative
vacuum is $\sim |F_S|^2\sim \cond^2/\mp^2= \lc^6/\mp^2$ (this is common
to the discussion in all the works referred to);
and (2) the form of the effective potential as one leaves
$\lm=0$ is determined by the largest field-dependent mass, which
was shown in Ref.~\cite{ross} to be that of the (hidden) gluino:
\be
m_{\tilde g}(\phi)\sim \left(\frac{\mp}{\lc}\right)\ \ms(\phi) \sim
\frac{\lm}{\lc\mp}\sim \frac{\phi^3}{\lc\mp}
\label{eq:mgluino}
\ee
so that \cite{ross}
\be
V(\phi)\sim \frac{\lc^6}{\mp^2}-\lc^4\left(\frac{m_{\tilde g}}{\lc}\right)^2
= a\frac{\lc^6}{\mp^2}-b\frac{\phi^6}{\mp^2}+\ldots
\label{eq:vphi}
\ee
with the additional parts of the potential assuring that $V$ attains its
minimum at $\phi\sim \lc.$ The same form of the downward slope $(\sim
(\lm)^2/\mp^2)$ is observed in the last term
of the potential given in Ref.~\cite{taylor}, where it is generated by the
inclusion of moduli.

Where is the dilaton in all this? The assumption here is that the roll of the
gaugino condensate to its final value $\lc^3$ takes place in a dilaton
background which, with the condensate small and in the slow-roll phase, is
roughly constant. Its presence is subsumed in the constants $(a,b).$
Eventually, as the condensate begins its fast motion toward
$\lc^3,$ the dilaton will move to its own stationary value.
Of course, we have no definitive idea of
what stabilizes the dilaton, but the quadratic behavior of the dilaton
potential near its stationary point is of crucial importance to the question
of relics, and will be addressed later in this work. At any rate, the
hypothesis to be tested at this stage is whether a successful inflationary
physics can be realized through $\phi$ as an inflaton, with $V$ in
Eq.~(\ref{eq:vphi}) as
the candidate for a ``New Inflation'' type of potential \cite{models}.
\bigskip

\noindent{\large\bf Destabilization of $\lm=0.$}\medskip

As the temperature decreases to $\lc$ (where the hidden gauge sector forces
become
strong), the effective potential at one loop will be dominated by the
(field-dependent) effective mass of the hidden gluino, $m_{\tilde g}\simeq
\phi^3/(\mp\lc)$ \cite{ross}. This then modifies Eq.~(\ref{eq:vphi})
to read, at high
temperatures,
\be
V_{eff}(\phi,T)\simeq V(\phi)+ T^2m_{\tilde g}^2\simeq
a\frac{\lc^6}{\mp^2}-b\frac{\phi^6}{\mp^2}+b'\frac{T^2}{\lc^2}
\frac{\phi^6}{\mp^2} + \ldots \  \ .
\label{eq:vphit}
\ee
{}From (\ref{eq:vphit}), it is then apparent
that the destabilization of the $\lm=0$ vacuum indeed begin to take
place at $T\simeq \lc.$\clearpage

\noindent{\large\bf Onset of Inflation}\medskip

The era $T\simeq \lc$ does
not, however, define the beginning of the inflationary era, since the thermal
density at this time is much greater than the false vacuum energy: $\lc^4\gg
\rho_{vac}\simeq \lc^6/\mp^2.$ The inflationary era thus begins at a
temperature $T_i$ determined by
\begin{eqnarray}
\rho_{rad}&=&\rho_{vac}\nonumber\\
\mbox{or}\;\  \tin&\simeq& \left(\frac{30a}{g_*\pi^2}\right)^{\quarter}\
\lc\lcmp\  \ ,
\label{eq:tinf}
\end{eqnarray}
where $g_*\simeq 106.$ As the universe, in its radiation dominated phase, cools
from $\lc$ to $\tin,$ modes of $\phi$ with $k>\tin$ will be redshifted away. At
$T=\tin,$ $\phi$ will contain modes with $k\leq \tin,$ so that fluctuations of
$\phi$ about $\phi=0$ will be given by

\be
\delta\phi_{thermal}\simeq \sqrt{\frac{1}{12\eta}}\ \tin\simeq
\ \left(\frac{a}{5000\eta^2}\right)^{\quarter}\lc\lcmp\ \ .
\label{eq:phith}
\ee

I now discuss the conditions for slow-roll, and for  attaining a
sufficient number of $e$--folds of
inflation .
\bigskip

\noindent{\large\bf Slow Roll}\medskip

 This requires \cite{steinharturner} that
$V\prii<9\eta H^2,$ where $H$ is the Hubble constant during the
inflationary era. From Eq.~(\ref{eq:vphi}),
\be
H=\sqrt{\frac{8\pi}{3}}\frac{\rho_{vac}}{\mp}=\sqrt{\frac{8\pi
a}{3}}\frac{\lc^3}{\mp^2}\sim \ms\ \ ,
\label{eq:hi}
\ee
so that  slow roll can be maintained as long as
\be
\phi<\phi_{end}=\left(\frac{3\eta}{10b}\right)^{\quarter}(H\mp)^{\half}=
\left(\frac{4\pi}{5}\frac{a\eta}{b}\right)^{\quarter}\lc\lcmp
\ \
\label{eq:phiend}
\ee
where $V\prii(\phi_{end})=9H^2 \cite{albran}.$
\bigskip

\noindent{\large\bf Sufficient Inflation}\medskip

 For a scale crossing outside the
horizon when the field has value $\phi,$ the number of
$e$--folds of inflation is

\be
N(\phi)=3H^2\eta\int_{\phi}^{\phi_{end}}\frac{d\phi}{V\pri(\phi)}=
\left(\frac{\eta}{8b}\right)\left(H\mp\right)^2
\left(\frac{1}{\phi^4}-\frac{1}{\phi_{end}^4}\right)\ \ .
\label{eq:nphi}
\ee
In order to attain at least 50 $e$--folds of inflation, beginning values of
$\phi$ must satisfy
\begin{eqnarray}
\phi_{begin} \le \phi_{50}& =
&\left(\frac{\eta}{400b}\right)^{\quarter}(H\mp)^{\half}
=\left(\frac{\pi}{150}\frac{\eta a}{b}\right)^{\quarter}\lc\lcmp\nonumber\\
\mbox{or}\;\ \phi_{begin}&\le &(120)^{-{\quarter}}\phi_{end}\ \ .
\label{eq:phibeg}
\end{eqnarray}
We note that $\phif$
is of the same order $(\sim\lc\lcmpl)$ as $\delta\phi_{thermal}$, so that
requiring $\phif\geq\delta\phi_{thermal}$ imposes (from Eqs.~(\ref{eq:phith})
and (\ref{eq:phibeg})) a condition on the parameters $b$ and $\eta:$
\be
\etab\leq 3\ \ .
\label{eq:etab}
\ee
\bigskip

\noindent{\large\bf Density Fluctuations}\medskip

Following the usual prescriptions \cite{drr},
scalar density fluctuations on a scale $\lambda$ re-entering
the horizon during the matter-dominated era may be calculated via
\be
\drr = \frac{2}{5}\ \frac{H\delta\phi}{\dot\phi(t_1)}=
\frac{1}{5\pi\sqrt{\eta}}\ \frac{H^2}{\dot\phi(t_1)}\ \ ,
\label{eq:drra}
\ee
with $\phi(t_1)$ the value of $\phi$ when the scale $\lambda$ first crosses
outside the horizon during inflation \cite{lambda}. Using the potential
(\ref{eq:vphi}) in the equation
of motion for $\phi,$ in the slow roll approximation, one may easily evaluate
this quantity in terms of $\nl$, the number of $e$--foldings of inflation
subsequent to the first horizon crossing of a scale $\lambda:$
\begin{eqnarray}
\drr&=&\frac{8}{5\pi}\ \left(\frac{b}{2\eta^3}\right)^{\quarter}\
\left(\frac{H}{\mp}\right)^{\half}\nl^{\frac{5}{4}}\nonumber\\
&=&\frac{8}{5\pi}\ \left(\frac{4\pi a b}{3\eta^3}\right)^{\quarter}\
\left(\frac{\lc}{\mp}\right)^{\frac{3}{2}}\ \nl^{\frac{5}{4}}\nonumber\\
&\simeq & \ac\ \etab\ \left(\frac{\ms}{\mp}\right)^{\half}\nl^{\frac{5}{4}}
\label{eq:drrb}
\end{eqnarray}
with the parameter $c$ defined by $\ms=c\lc^3/\mp^2\sim H.$ The $e$--folding
quantity $\nl$ is given by
\be
\nl\simeq 36+\ln
\lambda_{\mbox{Mpc}}+\tthird\ln(H/1\tev)+\tthird\ln(T_{RH}/1\tev)\ \ .
\label{eq:enlambda}
\ee
I will shortly demonstrate that
the reheat temperature $T_{RH}\sim 1\tev\sim H,$ so that for $\lambda$
in the
range $1\rightarrow 10^4$ Mpc, at least 45
$e$--foldings of expansion are required to solve the entropy problem.

An estimate of the first factor in (\ref{eq:drrb})  may be obtained from
the tree-level supergravity result, Eqs.~(26) and (27) in Ref.~\cite{ross}: for
an
$SU(N)$ gauge theory with $n_f$ fermions,
\be
\left(\frac{a}{c^2}\right)^{\quarter}=
\left(\frac{V_0}{\mp^2 m_{3/2}^2}\right)^{\quarter}=
\left(1+\frac{16\pi^2}{g^2\left(N-\tsixth
n_f\right)}\right)^{\half}\simeq 10\ \ .
\label{eq:ac}
\ee
On interpolating
$\nl$ about
$\lambda=100$ Mpc,  and inserting the estimate (\ref{eq:ac}), one obtains from
(\ref{eq:drrb}) the principal result of this work:
\begin{eqnarray}
\drr&\simeq& \etab\ \cdot 10^3
\left(\frac{\ms}{\mp}\right)^{\half}\left(\frac{\lambda}{100\
\mbox{Mpc}}\right)^{0.03}\nonumber\\
&\simeq&10^{-5}\left(\frac{\lambda}{100\
\mbox{Mpc}}\right)^{0.03}\left(\frac{b}{\eta^3}\right)^{\quarter}\ \ .
\label{eq:drrc}
\end{eqnarray}

The COBE data
\cite{cobe} imply
$(\delta\rho/\rho)_{hor,\lambda}\simeq 2\times 10^{-5}$
on scales of $\sim 10^3$ Mpc, so that the result (\ref{eq:drrc})  agrees
with COBE if
$\left(b/\eta^3\right)^{\quarter}\simeq 2.$ This is a plausible number,
consistent with the thermal constraint (\ref{eq:etab}). Optimistically, one may
regard this requirement on $\left(b/\eta^3\right)^{\quarter}$ as not in the
``fine-tuning'' category, and thus advertise the result (\ref{eq:drrc}) as an
agreement with COBE involving (essentially) {\em no adjustable parameters.}

This accomplished, I turn to the crucial question of the relic problem.
\bigskip

\noindent{\large\bf Relics and Reheating}\medskip

At the end of the slow roll, the
field rolls quickly to its minimum at $\phi\simeq \lc,$ and begins to
oscillate. The persistence of the energy of oscillation into the present era,
with an energy incompatible with the presently measured Hubble constant, or
of the decay of the particles late enough to cause post-nucleosynthesis
reheating, is  a well-known problem associated with inflation based on
``hidden'' sector
fields,  whose interactions are purely gravitational \cite{tevrelics}.
I wish to show how this
problem disappears in the present case.

I begin by sketching a calculation of the mass spectrum and relevant
interactions
of the  hidden relic sector \cite{gb}.  In the
conventional description, the
condensate  and
dilaton fields roll to the minimum of the $S$-$\lm$
potential \cite{dine2}. Field fluctuations about the minimum,
as calculated from $\partial^2
V/\partial \sigma^2,\ \partial^2 V/\partial G^2,$ consist of
glueballs $(G)$ of mass
$\sim \lc,$ and dilatons ($\sigma)$ of the universal gravitino mass
$\ms\sim m_{3/2}\sim \lc^3/\mp^2.$ Since the
$\sigma$ decays only through gravitation-
strength
interaction, it is easy to show that for $\ms\sim 1\tev,$ the
reheat temperature lies below $1\ $MeV, thus destroying the basis of the
nucleosynthesis calculations.  In the present case, there is an even more
exigent situation: since inflation occurs so late, it is important that
reheating take place at a temperature {\em above the electroweak scale}, if
there is to be any chance at all for leaving a baryon asymmetry.
What I will  now
show is that the residual $\sigma$-$G$ interaction in fact destabilizes the
naive vacuum and radically alters the spectrum, so that reheating
takes place at $1\tev.$

The $\sigma$-$G$
interaction is the result of the original superpotential (\ref{eq:w}). It
is of $O(1/\mp),$ and can be taken (phenomenologically) as
${\cal L}_{int}\sim (\lc^3/\mp)\ \sigma G.$ Introducing some quartic terms
for
stability, we may incorporate the above considerations into an
effective potential near $\sigma=G=0:$
\be
V_{eff}(\sigma, G)\simeq \thalf\mp^2\left(\alpha\epsilon^6\sigma^2+
\beta\epsilon^2 G^2+\kappa\epsilon^3 \sigma G\right)
+\tquarter\lambda_{\sigma}\sigma^4 +\tquarter\lambda_G G^4
\label{eq:sigmagpot}
\ee
where $\epsilon\equiv \lc/\mp,$ and $\alpha\sim\beta\sim|\kappa|\sim
O(1).$
A simple calculation then shows that the mixing term generates a
global minimum with energy density $\sim
O(\lc^8/\mp^4)$ below the (false) $\sigma=G=0$
vacuum. The shift of the dilaton
field
from $\sigma=0$  is of $O((\lc/\mp)^2),$ which is $\ll g^{-2}=S_0.$
However, a calculation of the fluctuations about  this  minimum yields a
revised mass spectrum
\begin{eqnarray}
\msigp&\simeq&\left(\frac{2\kappa^2}{\beta}\right)^{\half}
\frac{\lc^2}{\mp}\sim\left(\frac{\mp}{\lc}\right)\ms\nonumber\\
m_{G\pri}&\simeq&\sqrt{\beta}\lc\simeq m_G
\label{eq:msigmg}
\end{eqnarray}
Note that $\msigp$ has no dependence on the original $m_{\sigma}$, nor on
$\lambda_{\sigma}$ or $\lambda_G$ (as long as $\lambda_{\sigma}$ is non-zero).
In terms of the unmixed fields,
\be
\sigma\pri\simeq \sigma+O(\epsilon^2)G\ \ ,
\label{eq:sigpsig}
\ee
so that the coupling of $\sigma\pri$ to ordinary particles (say two gluons)
is of $O(\msigp^2/\mp),$ giving a decay width
\be
\gsigp\simeq \frac{\msigp^3}{\mp^2}\sim \frac{\lc^6}{\mp^5}\ \ .
\label{eq:gsigp}
\ee

It now turns out that {\em this precisely solves the relic problem.}
Since $\gsigp\ll
H\sim \lc^3/\mp^2,$ we have a situation of inefficient reheating. In the
usual fashion, we can calculate the reheat temperature to be
\be
T_{RH}=\left(\frac{15}{\pi^3g_*}\right)\left(\mp\gsigp\right)^{\half}\sim
\frac{\lc^3}{\mp^2}\sim 1\tev \ \ ,
\ee
so that there is no problem with either nucleosynthesis or electroweak
baryogenesis. By contrast, the reheat
temperature of a relic with $m=\ms$ is of
$O\left(\left(\frac{\ms}{\mp}\right)^{\half}\ \ms\right),$ which  requires
a very large value for $\ms$ in order to avoid low temperature reheating.
\bigskip

\noindent{\large\bf Summary and Conclusions}\medskip

\noindent (1)
I  have presented a dynamical model for inflation based on SUSY-breaking in a
hidden sector containing  dilaton and gauge superfields.
The identification of the inflaton as a scalar field whose cube is the
hidden gaugino condensate, with a potential suggested by previous work
\cite{taylor,ross}, has led to a version of inflation in which the
usual small coupling is replaced by the small dimensionless ration $\ms/\mp.$

\smallskip

\noindent(2) Inflation sets in relatively late (at $T\sim 10^{10}\ GeV).$
Reheating via decay of heavy dilatons (heavy by virtue of mixing with
glueballs)
is inefficient, and the reheat temperature is about
$1\tev.$ Thus,  baryogenesis must take place at temperatures below $1\tev,$
with the most likely possibility being the anomalous electroweak process
\cite{kuzmin}.\smallskip

\noindent(3) The dominant aspect of the density fluctuations generated by
deSitter fluctuations is largely determined by the small
ratio $(\ms/\mp)^{\half};$
perhaps a factor of 2 remains to  be decided by a
specific combination
$(b/\eta^3)^{\quarter}$ of  unknown parameters characterizing the
strongly-interacting phase of the dilaton-gaugino system. Thus, the qualitative
behavior of the density fluctuation is directly predictable from the
dynamics of the
SUSY--breaking sector.

\noindent(4) Finally, the present work, in common with much other in the field,
offers no clue with respect to solving the cosmological constant problem. The
potential (\ref{eq:vphi}) must be set to zero at its minimum by hand.

\section*{Acknowledgment}
I would like to acknowledge many helpful conversations with Tom Taylor
concerning recent work on string-based SUSY-breaking mechanism.
 This research was supported in part
by the National Science Foundation (Grant No. PHY-9306524), and by the
Texas National Research Laboratory Commission (Award No. RGFY93-277).


\begin{thebibliography}{99}
\bibitem{lambda} The comoving size of $\lambda$ is
normalized to its physical size today.
\bibitem{dine2} In what follows, I explicitly assume that a minimum of the
$S$--$\lambda\lambda$ potential has been established, probably through
non-perturbative effects \cite{dine}.
There are, of course many questions surrounding the dilaton
potential \cite{steinbru}, but perhaps a satisfactory solution can be
found in the context of a complete $S$--$T$--$\lambda\lambda$ potential.
\end{thebibliography}
\end{document}